\pdfoutput=1
\documentclass[letterpaper,twocolumn,10pt]{article}
\usepackage{usenix}
\usepackage{cite}
\usepackage{amsmath,amssymb,amsfonts}
\usepackage{algorithm,algorithmic}
\usepackage{graphicx}
\usepackage{textcomp}
\usepackage{xcolor}
\usepackage{hyperref}
\usepackage{cleveref}
\usepackage{booktabs}
\usepackage[justification=centering]{caption}
    
\begin{document}

\hypersetup{
    pdftitle={Automatic Bill of Materials},
    pdfauthor={Nicholas Boucher \& Ross Anderson}
}

\title{Automatic Bill of Materials}

\author{
{\rm Nicholas Boucher}\\
\small{University of Cambridge}\\
\small{\rm{}nicholas.boucher@cl.cam.ac.uk}
\and
{\rm Ross Anderson}\\
\small{Universities of Cambridge \& Edinburgh}\\
\small{\rm{}ross.anderson@cl.cam.ac.uk}
}

\renewcommand{\algorithmiccomment}[1]{\hfill$\rhd$#1}
\makeatletter
\renewcommand\@makefntext[1]{\leftskip=1.5em\hskip-.75em\@makefnmark\ #1}
\makeatother

\maketitle

\begin{abstract}
Ensuring the security of software supply chains requires reliable identification of upstream dependencies. We present the Automatic Bill of Materials, or ABOM, a technique for embedding dependency metadata in binaries at compile time. Rather than relying on developers to explicitly enumerate dependency names and versions, ABOM embeds a hash of each distinct input source code file into the binary emitted by a compiler. Hashes are stored in Compressed Bloom Filters, highly space-efficient probabilistic data structures, which enable querying for the presence of dependencies without the possibility of false negatives. If leveraged across the ecosystem, ABOMs provide a zero-touch, backwards-compatible, drop-in solution for fast supply chain attack detection in real-world, language-independent software.
\end{abstract}
\section{Introduction}

The complexity of modern software supply chains poses a significant threat to software security. The vast collection of prebuilt solutions to common software engineering tasks available in both open and closed source ecosystems encourages developers to leverage these implementations in downstream software. Such practices have many benefits ranging from decreasing software production costs to increasing the likelihood that implementations are crafted by domain experts. However, the resulting dependency graphs increase the blast radius of vulnerabilities in upstream dependencies. This effect results in an increased appeal for supply chain attacks: attacks which target upstream dependencies with the intention of exploiting the collection of downstream software.

Understanding dependency graphs is more challenging than it may seem on the surface~\cite{bi2023way,xia2023empirical}. While it may be possible to perform static source code analysis to detect imports representing first level dependencies, it is more complicated to determine the second-level dependencies held by those dependencies. This process becomes yet more complex the further you travel up the dependency chain, or if any pre-compiled closed source dependencies are present.

The traditional solution to dependency identification is the Software Bill of Materials, or SBOM. SBOMs are an enumeration of dependencies provided with software as a list of software name and version pairs. These lists may be prepared manually by developers or via tooling, but either implementation leaves open the same set of challenges in accurately identifying upstream software. Early research into SBOM accuracy is concerning; in one study of the Java ecosystem, multiple major SBOM tools correctly recalled less than half of the ground truth dependencies and none of the examined tools provided perfect accuracy~\cite{balliu2023challenges}. Despite the need for improved techniques, SBOMs are quickly gaining adoption across the industry and are now required for suppliers to the US Government following Executive Order 14028~\cite{biden_eo14028_2021}.

We believe that a better solution for for dependency identification exists. An ideal solution is one that requires no effort to retrofit onto existing systems, provides a high level of accuracy, and is efficient in both space and time. We posit that any solution which requires developer action to guarantee accuracy is likely to fail; within the context of large dependency graphs it is likely that at least one developer will not take additional action beyond the minimum viable product.

We therefore propose the Automated Bill of Materials, or ABOM, a compile-time technique to embed highly space efficient metadata in binaries to support supply chain attack mitigation. We offer the key insights that such mitigation does not depend upon explicit dependency enumeration, but rather on the ability to test for the presence of specific known-vulnerable dependencies. Deployment of this technique across the ecosystem via compiler defaults will bolster supply chain defense by enabling easy, rapid detection of known-vulnerable code in downstream software.

In this paper, we make the following contributions:
\begin{itemize}
    \item We describe a novel technique for embedding source code hashes in binaries at compile time to support the fast identification of known-comprised code in downstream dependencies.
    \item We analyze a variation of Bloom Filters as a highly space efficient data structure for representing dependencies and conduct experiments to select optimal parameters for its construction.
    \item We implement the techniques described and demonstrate application to real-world software.
\end{itemize}
\section{Background}

In this section, we describe the prerequisite concepts that will be used to build a novel technique to identify supply chain attacks in software.

\subsection{Supply Chain Attacks}

Supply chain attacks are those in which an adversary attempts to inject malicious functionality into upstream dependencies that are leveraged by multiple downstream software products~\cite{5718996}. The victim downstream software may be operating systems, applications, or further shared software components. Supply chain attacks can be particularly appealing to adversaries because a single successful attack can lead to the simultaneous compromise of many different targets. These vulnerabilities are likely to persist within the ecosystem long after patches have been released~\cite{7163055}. Supply chain attacks are included in OWASP's top 10 web application security risks~\cite{owasp_a62021}.

\begin{figure}[t]
    \centering
    \includegraphics[width=\linewidth]{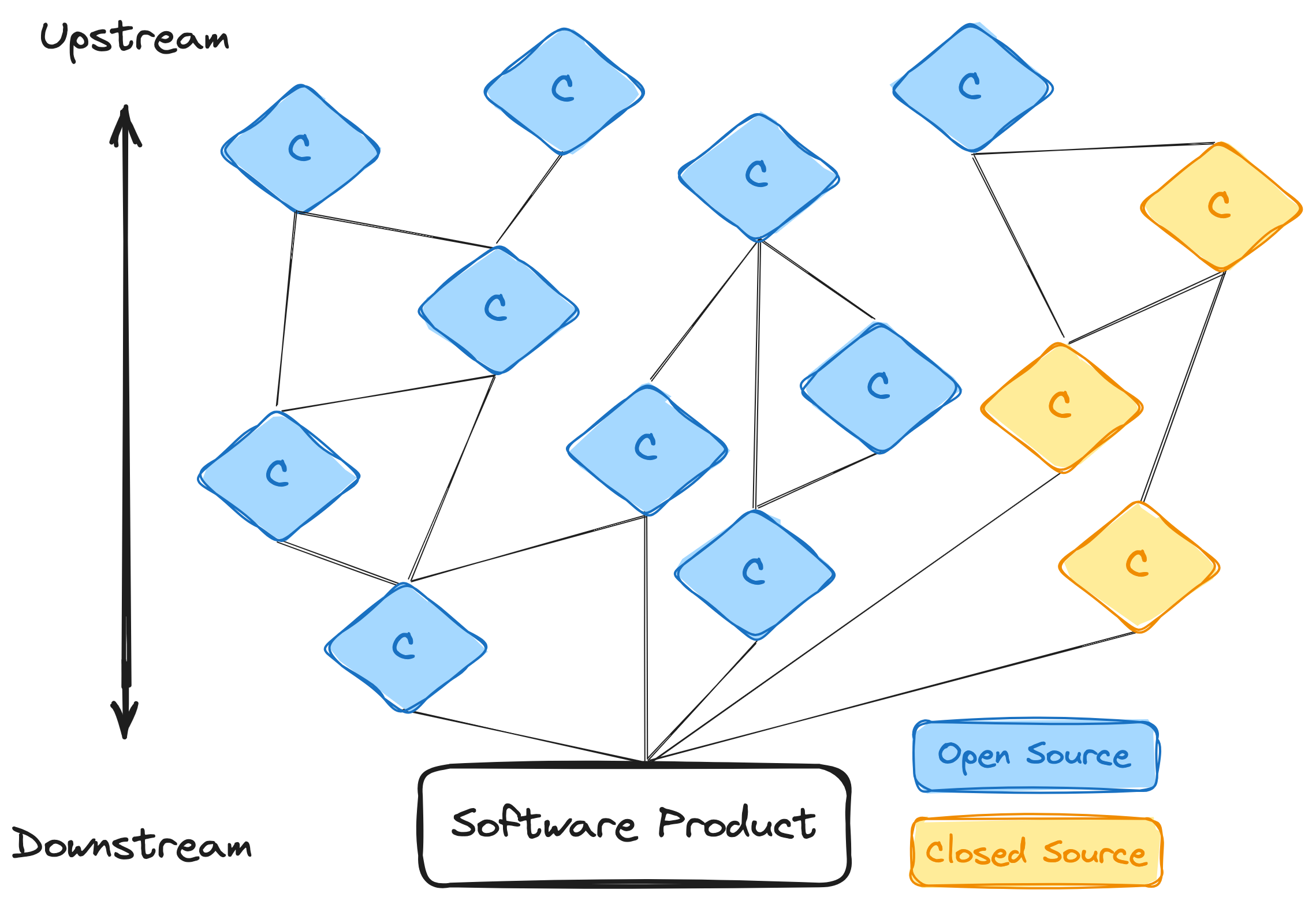}
    \caption{A software supply chain consisting of 13 upstream dependencies $C$ contributing to one downstream software product.}
    \label{fig:supply-chain}
\end{figure}

Modeling supply chains can be very challenging. Dependencies themselves take on dependencies, and the recursive depth of dependency chains can be arbitrarily large. We depict an example supply chain diagram for reference in \Cref{fig:supply-chain}. Social factors ranging from contracts to geopolitics can also play a meaningful role in dependency strategies~\cite{7180277}. Open source software, given its nature of public contributions, can be a powerful supply chain attack vector~\cite{10.1007/978-3-030-52683-2_2}. Code review partially mitigates this, but can still be subverted using underhanded techniques~\cite{boucher_trojansource_2023,wheeler_2020_underhanded}.

Supply chain attacks have existed in practice for multiple decades~\cite{1203227}, although the recent impact of significant attacks such as the Solar Winds incident~\cite{9382367} and Log4j incident~\cite{hnsw-rvmls-22} have drawn renewed attention.

\subsection{Software Bill of Materials}
\label{sbom}

Software Bills of Materials, or SBOMs, are lists of utilized components distributed with software~\cite{ntia_sbom_2021}. These lists should ideally be machine-readable, and a variety of formatting standards have emerged including SPDX~\cite{spdxspec_2022} and CycloneDX~\cite{cyclonedx_2023}. Commercial and open source tooling has emerged to help developers build SBOMs, although these tools often rely on package management systems to properly enumerate upstream dependencies. SBOMs are a key tool in mitigating supply chain attacks, as they enable software consumers to determine whether a vulnerable upstream component is in use.

Existing SBOM tools and formats support a wide variety of optional data fields for describing each software component, but the US National Telecommunications and Information Administration defines the following seven fields as the required components of an Executive Order 14028~\cite{biden_eo14028_2021} mandated SBOM~\cite{ntia_minsbom_2021}:
\begin{itemize}
    \item Supplier Name
    \item Component Name
    \item Version of the Component
    \item Other Unique Identifiers
    \item Dependency Relationship
    \item Author of SBOM Data
    \item Timestamp
\end{itemize}

\subsection{Bloom Filters}

\begin{figure}[t]
    \centering
    \includegraphics[width=\linewidth]{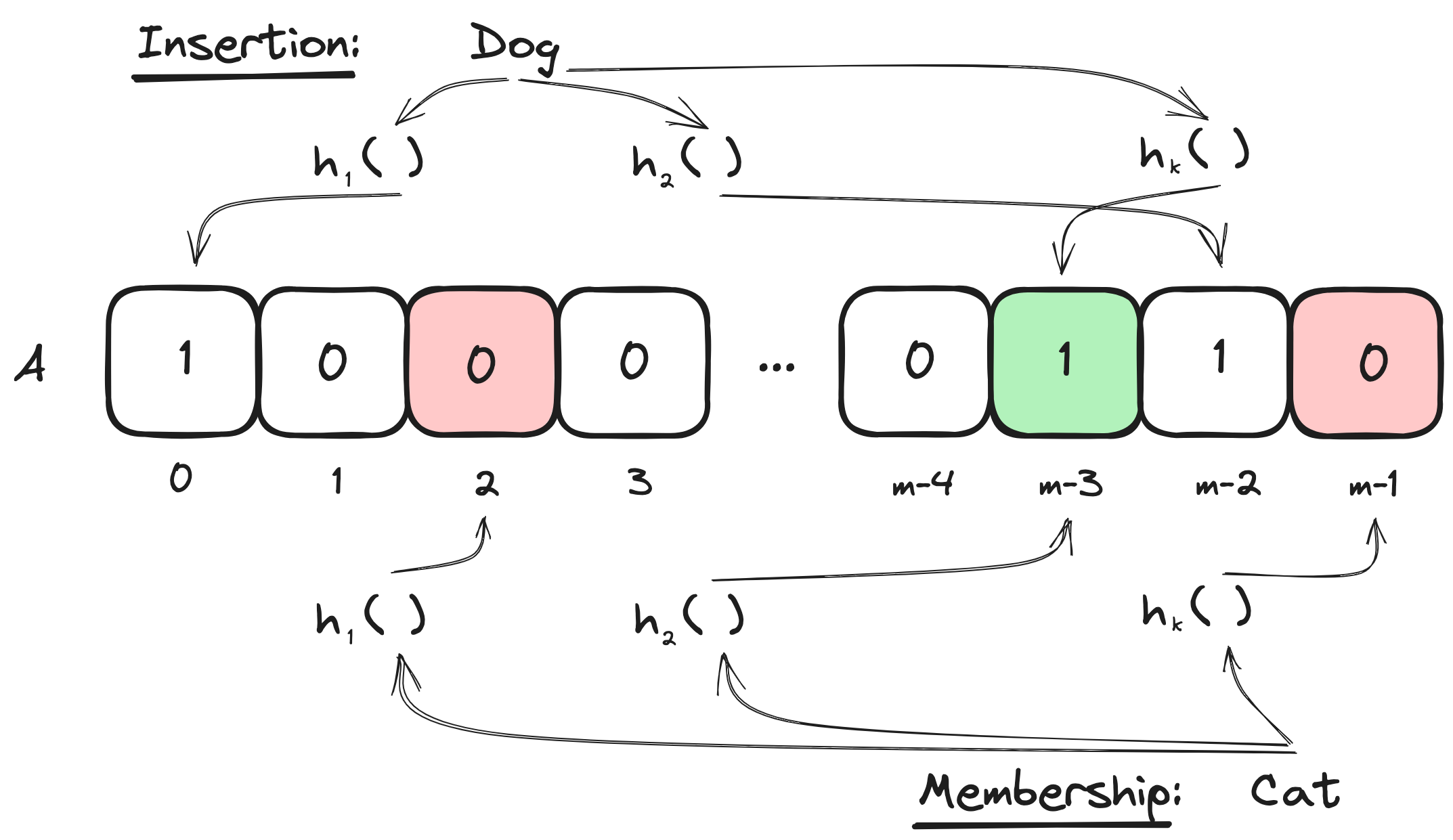}
    \caption{A Bloom filter in depicting the insertion of item \textit{Dog} and the failed membership query of item \textit{Cat}.}
    \label{fig:bloom}
\end{figure}

Bloom filters are highly space efficient probabilistic data structures for representing sets~\cite{bloom_1970}. They support insertions, unions, and membership queries. Membership queries have a tunable false positive rate, but false negatives are not possible; i.e. the filter may assert that it contains an item that it does not contain, but it will never assert that it does not contain an item that it does contain.

Bloom filters work as follows: during initialization, an array $A$ of size $m$ is initialized to all zeroes. When an element is inserted, it is hashed by $k$ independent hash functions to produce $k$ indices into $A$. The bits at each of these indices are then set to one. To query for set membership, the queried item is similarly hashed $k$ times and the resulting indices are checked. If all indices are set to one, the filter returns that the item is present and otherwise returns that the item is absent. Set unions can be performed by taking the bitwise OR of filter arrays. In classical Bloom filters, items cannot be removed once inserted. \Cref{fig:bloom} depicts Bloom filter insertion and queries.

Bloom filters have multiple tunable parameters: the array size $m$, the number of hash functions $k$, and the false positive rate $f$ when $n$ elements have been inserted. With careful parameter selection, Bloom filters can also be compressed for additional space efficiency at rest~\cite{mitz_compressed_2001}, although this introduces another parameter of compressed size $z$ when $n$ elements have been inserted. The specific hash functions chosen are also parameters of the filter; they must output values in the range $0\leq{}x<m$, but can be $k$ non-overlapping slices of suitably strong longer hash functions such as those of the SHA family.

As we will soon see, Compressed Bloom filters will enable us to construct an elegant, space-efficient alternative to SBOMs to help mitigate supply chain attacks.
\section{Design}

\begin{figure*}[t]
    \centering
    \includegraphics[width=.8\linewidth]{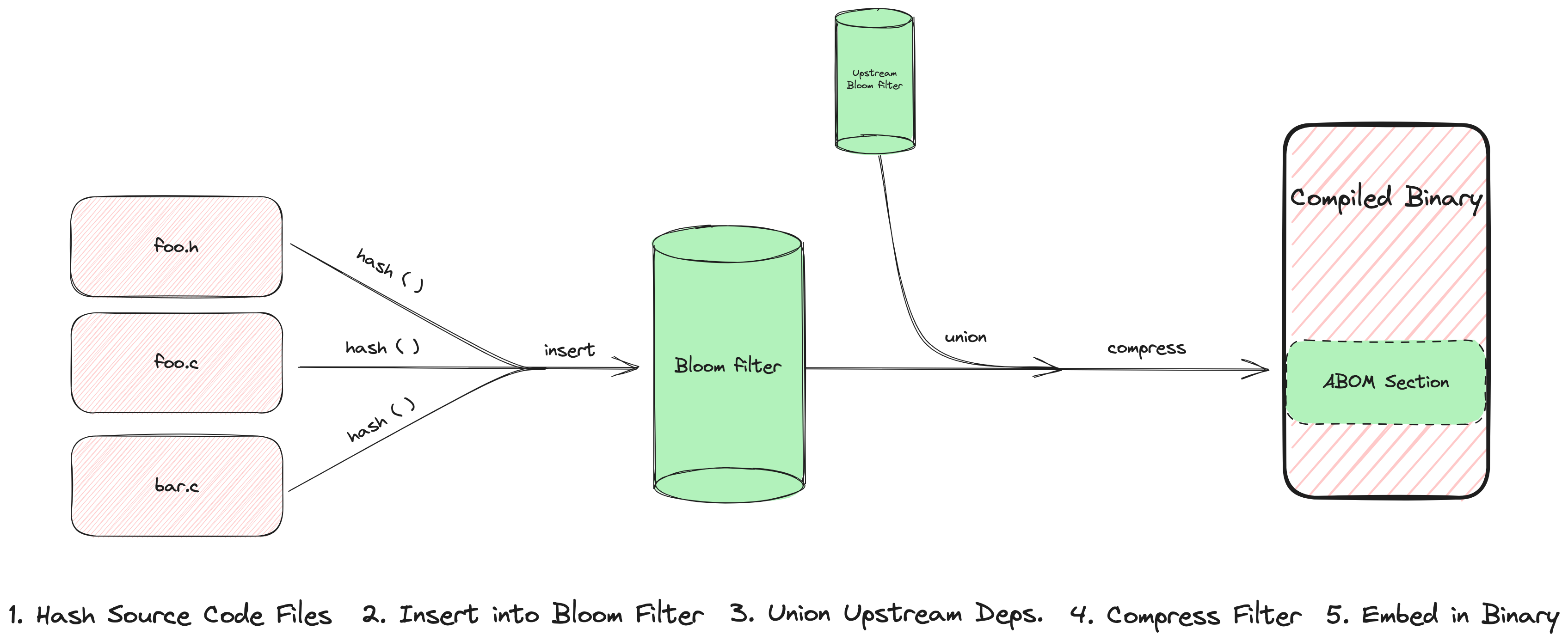}
    \caption{A visualization of the ABOM construction pipeline.}
    \label{fig:abom}
\end{figure*}

In this section, we propose a design for a highly efficient technique to assist with supply chain attack mitigation by identifying compromised software.

\subsection{Software Representation}

While a significant portion of modern software uses a versioning system such as Semantic Versioning~\cite{semver_2023} to identity different releases, not all software is versioned. This is particularly true for open source software and less mature projects. Furthermore, even versioned software will have development branches of the code base between versions, and these branches will themselves be functionally unversioned.

The lack of universality and precision in software versioning therefore creates a challenge for any systems that rely on it for vulnerability detection. From this, we observe that software supply chain attack identification techniques cannot rely exclusively on versioned software identifiers to precisely identify dependencies.

Instead, we propose a more robust mechanism for unambiguously identifying software independent of versioning practices: we represent software as the set of hashes of all source code files constituting that software.

We note that there is some room for ambiguity when selecting which source code files constitute software. For example, should C-style header files be included? What about graphics files referenced by source code? We suggest that any build input file which is transformed by a compiler into a different form as output should be included in the list of source code files. Following this logic, C-style header files should be included in the list of source code file hashes. This is sensible, as C-style header files can contain arbitrary C code even if that is not common practice. Graphics files, on the other hand, should not be included in the list of source code hashes \textit{unless} they are transformed by the compiler to package within the build output (which is not typically the case). This same logic should be extended to other languages and data types.

\subsection{Minimum Viable Mitigation}

There are many pieces of information that may be interesting when assessing the impact of a supply chain attack. However, we seek to identify the minimum amount of information necessary to mitigate a supply chain attack from the perspective of potentially compromised downstream software.

The minimum information needed determine whether a piece of software is impacted by a supply chain attack is whether a certain infected dependency was included somewhere along the supply chain. At first glance, it would therefore appear that the minimum information needed to mitigate a supply chain attack would be the list of all upstream dependencies used by a software product.

We can, however, do better. The key insight is that supply chain attack mitigation does not actually depend on enumerating all upstream dependencies, but rather on the ability to query whether a specific piece of compromised software was included as a dependency.

\subsection{Data Structure Selection}

We have already encountered a space-efficient data structure that implements queryable set representations: Bloom filters. By storing source code file hash sets in a Bloom filter, we will gain significant space savings compared to storing hashes directly. We will also gain constant-time hash queries.

Conveniently, since the items we are inserting into the Bloom filter are themselves hashes, we need not further hash inputs as part of the filter insertion routine. As long as the input hash $x$ is sufficiently large, we can simply leverage $k$ slices of length $log_2(m)$ bits to serve as $h_i(x)$ for $1\leq{}i\leq{}k$. This provides a significant time optimization for both insertion and membership querying, as further hashing is no longer required for data structure operation.

When incorporating a dependency for which source code hashes have already been inserted in a Bloom filter, these hashes can be added to the Bloom filter by simply taking the union with the upstream Bloom filter.

The drawback to Bloom filters is their probabilistic nature: there is some probability of a false positive when querying for the inclusion of a hash. Fortunately, we can tune this false positive rate to be very unlikely. In addition, it is key to note that there will never be false negatives: if a hash was inserted into the filter, there is no risk that the data structure ``forgets'' that hash.

In addition to the data structure parameters, the false positive rate of a Bloom filter is dependent on the number of distinct items that have been inserted. As the number of inserted items increases, so too does the false positive rate. However, while we want to limit the false positive rate we do not want to limit the maximum number of source code hashes that can be inserted into the data structure. Therefore, when the false positive rate reaches a set threshold, we opt to create an new Bloom filter for additional insertions. This is modeled on the design of Scalable Bloom Filters~\cite{ALMEIDA2007255}, but purposefully sets the $m,k$, and maximum $n$ parameters for each Bloom filter to be equivalent to allow for simple insertions, unions, and queries regardless of the number of filters present. When later selecting $f$, we will therefore account for the actual false positive rate being dependent upon the number of Bloom filters present.

In the unlikely scenario where a Bloom filter reports a false positive for a vulnerable dependency, the list of hashes constituting the Bloom filter can be produced rapidly as a witness to invalidate the false positive.

\subsection{Compression}

If the number of source code hashes inserted into a Bloom filter is small, it is possible for the Bloom filter to increase the space required to represent dependencies rather than decreasing it as desired. In this scenario, we observe that the majority of the Bloom filter will be repeated zeroes, and therefore the data structure would be a good candidate for compression. However, this compression benefit would be lost as the number of items inserted into the Bloom filter grows.

Instead of using traditional Bloom filters, we can use Compressed Bloom filters~\cite{mitz_compressed_2001}. This variation of Bloom filter selects parameters such that the data structure is a good candidate for compression when stored. Typically, this takes the form of selecting large $m$ and small $k$ for a given $f$. Selected appropriately, the compressed size $z$ is smaller than $m$ would have been if it was optimized for the same $f$ without compression.

By using a Compressed Bloom filter to store source code hashes, we arrive at a solution that is highly space efficient for both large and small collections of source code files.

\subsection{Packaging}

The techniques thus far described are only effective if they are performed for every upstream dependency. When a dependency is compiled, the downstream user no longer has access to its source code, and is therefore reliant on the upstream publisher to distribute the Compressed Bloom filter of hashes for this scheme to succeed.

We do not expect that any system that requires software publishers to distribute auxiliary files along with compiled binaries will be adopted ubiquitously. Compiled binaries are the minimum product that must be distributed to ship software, and it is reasonable to expect that at least some software producers will ship only this.

Therefore, we propose that the best place to ship bills of materials, including our Compressed Bloom filters, is embedded within compiled binaries. All major executable formats, including Linux's ELF, MacOS's MachO, and Windows' PE, support the ability to embed arbitrary non-code data as named binary sections. It thus follows that our source-code-hash-containing Compressed Bloom Filters be embedded as a dedicated section in compiled binaries.

From this, it becomes apparent that build time is the natural stage in which to generate these binary dependency sections. An ideal solution is one that is built into the compiler and enabled by default. In this setting, such bills of material would quickly become ubiquitous across the ecosystem, as they would require zero touch for developers to construct and produce no additional artifacts that need to be shipped with together with compiled binaries.

\begin{figure*}[t]
    \centering
    \includegraphics[width=.8\linewidth]{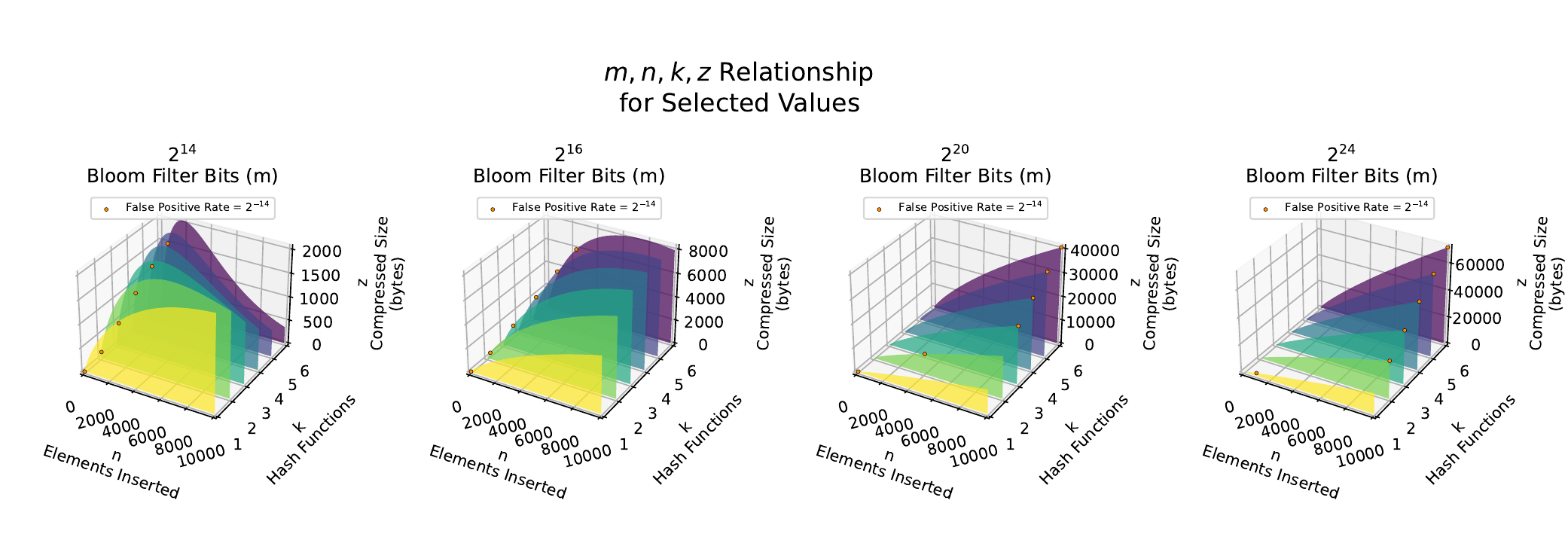}
    \caption{Visualizing the relationship between $m,n,k,z$ to assist with selecting optimal values.}
    \label{fig:combined-4d}
\end{figure*}

\subsection{ABOM}

Collectively, we refer to the techniques described in this section as the Automatic Bill of Materials, or ABOM.

In summary, ABOM represents software dependencies as the set of hashes of all source code files later ingested by a compiler. These hashes are then stored in a Bloom filter which is compressed when written to disk. Bloom filters from upstream dependencies are merged downstream via Bloom filter unions at build time. These data structures are packaged into compiled binaries as named sections within the executable files emitted by compilers. The pipeline is visualized in \Cref{fig:abom}.

In the event of a supply chain attack, ABOM would be used to assist mitigation as follows: first, after the attack is discovered, the upstream dependency producer calculates the hashes of all versions of the source code file containing the vulnerability and publishes these values. Downstream software users then extract the ABOM from the compiled product binary, decompress the Bloom filter, and query for the presence of a known-compromised hash. If the Bloom filter returns a match, then the software is considered infected. We note that this downstream ABOM querying would likely be automated by antivirus tooling.
\section{Parameter Selection}

There are a variety of parameters that must be selected to implement ABOM as described in the previous section. In this section, we will discuss, optimize, and select these parameters.

\subsection{Hash Function}

ABOMs represent dependencies as collections of source code hashes. It is therefore necessary to select a hash function for this purpose.

Our criteria for selecting a hash function are similar to the general criteria: we want preimage resistance to preserve the confidentiality of non-public source code files, efficient computational performance, and minimal collisions so that it can be reasonably modeled as an ideal hash function. One additional consideration is that we would like to select a hash function that is widely implemented within existing developer tooling. The reason for this is that we want to create the lowest possible barrier for a developer to calculate and publish the hash of a known-compromised file in the event of a supply chain attack.

The SHA family of hash functions meets these criteria. Within this family we omit SHA-1 due to known attacks~\cite{10.1007/11535218_2} leaving the selection between SHA2~\cite{sha2-fips180} and SHA3~\cite{sha3_2015}. While SHA2 is more widely used in the current ecosystem, we select SHA3 in anticipation of its overtaking SHA2 in prevalence following the publication of FIPS-202~\cite{sha3-fips202}.

Within the SHA3 suite we will select the hash bit length of choice according to the number of bits needed to index into the Bloom filter. We will calculate this in the following section.

\subsection{Bloom Filter Configuration}

There are a collection of parameters that define Bloom filters: the length $m$ of the array $A$, the number of hash functions $k$, and false positive rate $f$ when $n$ elements have been inserted. From the original Bloom filter proposal~\cite{bloom_1970}, we know that the relationship between the parameters is:
\begin{equation}
    f = \biggl(1 - \Bigl(1 - \frac{1}{m}\Bigl)^{kn}\biggl)^k
\end{equation}

However, the later Compressed Bloom filter proposal~\cite{mitz_compressed_2001} adds a fourth parameter: the compressed size $z$ of the filter. While the actual compressed size will depend on the compression algorithm selected, we follow from earlier results to model $z$ according to an optimal compressor with output determined by the entropy function:
\begin{equation}
\label{alg:entropy}
    z = m\bigl(-p\log_{2}p - (1 - p) \log_2(1 - p)\bigl)
\end{equation}

where $p$ is defined:
$$ p = \Bigl(1 - \frac{1}{m}\Bigl)^{kn} $$

Using these relations, we seek to select optimal values of $m,k,n,f,z$ for the purposes of the ABOM.

First, we opt to fix the maximum false positive rate $f$ that we are willing to tolerate. Clearly, we would like false positives to be rare, but how should we quantify rarity? We opt to answer this subjective question by mapping to something else the authors find rare: the lifetime odds of someone being struck by lightning in the US are $1/15300$~\cite{nws_lightening_2018}. We select $2^{-14}$ -- the next smallest power of 2 -- as our maximum value for $f$. Heuristically, an IT department scanning hundreds of critical systems against dozens of possibly relevant CVEs per month might have a handful of false positives a year; enough to exercise the system but not enough to swamp the true positives and lead to decreased vigilance.

We further restrict that $m$ must be a power of 2 to provide the convenient feature that an index can be represented as any $\log_2m$ length bit sequence. Finally, we observe that the optimal $k$ value will be a small, likely single-digit binary number due to the properties exuded by Compressed Bloom filters~\cite{mitz_compressed_2001}.

With these constraints established, we begin to seek the optimal set of parameters for $m,n,k,f,z$. For these purposes, optimal parameters will be those that meet the constraints already outlined while jointly minimizing $z$, maximizing $n$, and ensuring that $m$ is sufficiently sized to fit comfortably in the memory of average commercial hardware.

\begin{figure*}[t]
    \centering
    \includegraphics[width=.8\linewidth]{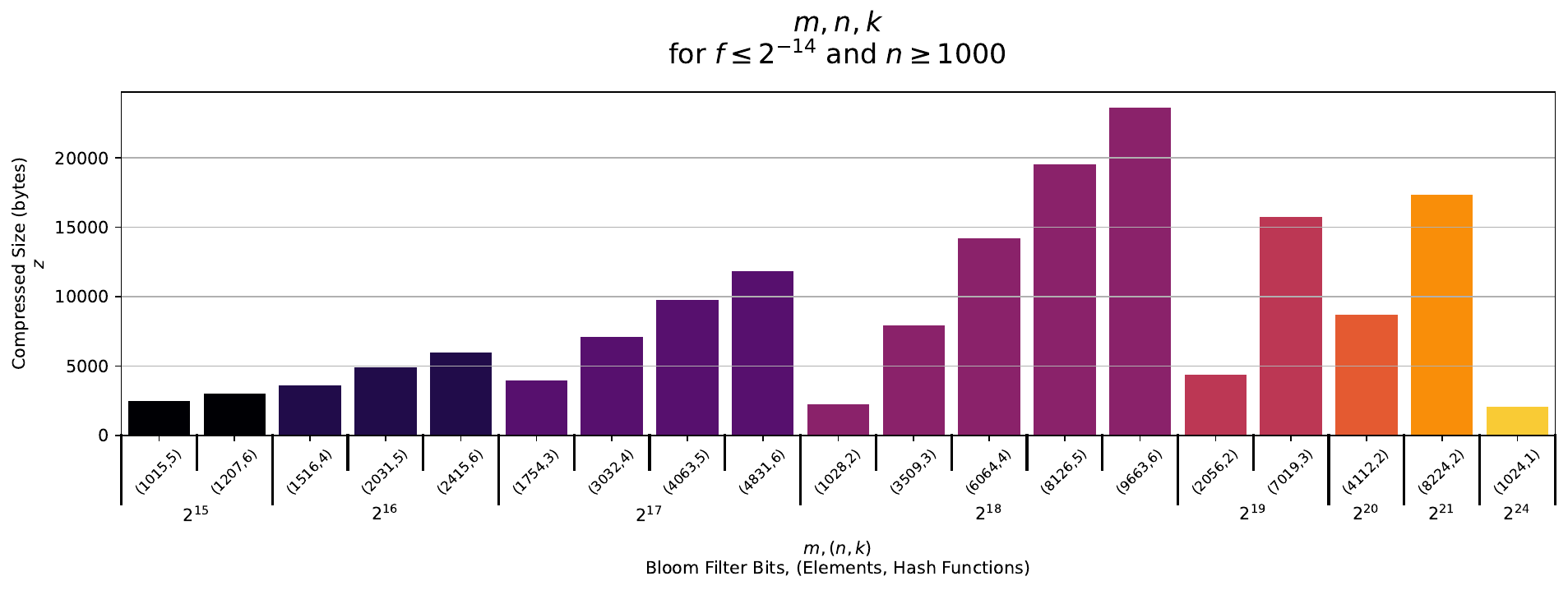}
    \caption{All $m,n,k$ plotted by $z$ where $m\leq2^{24}, f\leq2^{-14}, n\geq1000$ .}
    \label{fig:optimals}
\end{figure*}

To better understand the relationship between these parameters, we plot $n,k,z$ for four different values of $m$ that satisfy our constraints in \Cref{fig:combined-4d}. We also note the points at which $f$ reaches $2^{-14}$ on each plot. A variety of trends becomes clear: (1) $z$ grows with $k$, (2) $f$ grows with $n$ at a rate inversely proportional to $k$, and (3) $f$ shrinks with both increasing $m$ and $k$. We also note that $z$ has a negative parabolic relationship with $n$, which is derived from the fact that the compressed size is smallest when the filter is either all zeroes or all ones.

Using these visualizations, we next seek to select a constraint on $n$, which we hope will combine with the existing constraints to significantly narrow the set of potential parameter combinations. We select that $n\geq1000$. 1,000 is selected both because we posit that an average piece of software likely contains less than 1,000 contributing source code files, but also because selecting this constraints results in a relatively small number of remaining constraint-satisfying parameter permutations.

We depict all possible parameter combinations that meet our final set of constraints in \Cref{fig:optimals}. From this plot, we can see that there are now only 19 possible parameter permutations that satisfy the constraints thus far described. From these parameter combinations, we prefer those that have the smallest compressed size $z$.

We therefore list the five parameter combinations from the 19 options which have the lowest $z$ values in \Cref{tbl:optimals}. These will be the final options from which we select our optimized parameters.

Given all constraints, the options which results in the smallest compressed size are $m=2^{24}, k=1$. However, we note that this value of $m$ is somewhat large occupying over 2MB of contiguous space in memory. During compilation, it is likely that multiple Bloom filters will need to be held in memory, e.g. when linking multiple objects. The memory required could therefore grow quickly, and while many machines would easily be able to support these space requirements we would prefer to select parameters that would not limit the possibility of compiling/validating ABOMs on lower memory machines.

We therefore select the constraint-satisfying parameter combination with the second smallest compressed size as our chosen parameters. This parameter set of $m=2^{18}, k=2$ requires less than 33kB of memory to hold an ABOM, which is a space requirement that can be easily met by entry-level modern hardware. When the filter is saturated to 1028 items inserted, this results in an expected compressed size of 2160 bytes (2.16kB) requiring just an expected 2.1 bytes to represent each source code file hash!

To enable scenarios with large numbers of dependencies, ABOM supports multiple Bloom filters packaged in sequence. When this occurs, the false positive rate compounds with the increased number of filters. The cumulative false positive rate $\bar{f}$ for $q$ successive filters is therefore:
\begin{equation}
\bar{f} = 1 - (1 - f)^q
\end{equation}
While it is possible to decrease the false positive rate of subsequent Bloom filters to prevent the cumulative false positive rate from growing~\cite{ALMEIDA2007255}, we opt not to take that approach. If we were to select different parameters for subsequent Bloom filters to bound the cumulative false positive rate, this would mean that unions could not be taken arbitrarily with other Bloom filters of different configurations -- a process necessary for linking the ABOMs of different objects. Even if we were to limit the  max number of elements $n$ inserted into subsequent Bloom filters to bound the $\bar{f}$, this too would prevent building an ABOM that links together more than one saturated Bloom filter. Consequently, we choose to keep the same parameters for all sequential Bloom filters packaged in an ABOM and accept that $\bar{f}\geq{}f$. This effect motivated the selection of the max $f$ value tolerance above, which was rounded down to a smaller value than the target error tolerance to offset the effect of $\bar{f}$.

When constructing ABOMs, we will create a new Bloom filter whenever an insertion or union would cause $n$ to grow above our selected parameter  of 1028, which will in turn keep $f$ within our $2^{-14}$ bound. However, we note that the true value of $n$ will not be known in practice as Bloom filters do not directly track the number of elements inserted. Keeping a counter for the number of elements inserted into each Bloom filter would not only increase the amount of storage space required, but would also require handling the counting of duplicate insertions. Instead, we build on the observation that the expected value of $n$, which we will denote $n^*$ can be estimated by the number of ones $x$ present in the filter~\cite{doi:10.1021/ci600526a}:
\begin{equation}
    n^* = \mathbb{E}(n|x) = -\frac{m}{k}\ln\left(1-\frac{x}{m}\right)
\end{equation}

Using this value, we can approximate when $n$ reaches 1028, and use this as a trigger to generate additional Bloom filters of $m=2^{18},k=2$.

Since the elements inserted into ABOM Bloom filters will themselves be hashes, it is unnecessary to further hash inputs for index generation. We need a deterministic method for generating hashes with the correct number of bits needed. Until the appearance of FIPS 202, the canonical way to do this would have been to hash the input using (say) SHA-256 and then expand this to the desired length using (say) AES in counter mode. Thankfully, FIPS 202 provides the extendable-output function SHAKE~\cite{sha3-fips202} within the SHA3 family which does the work for us, and which will likely be commonly available in developer tooling.

For each index into the Bloom filter, we will need an index of $\log_2(m)$ bits. For our chosen value of $m=2^{18}$, this is 18 bits. We have also selected that there will be $k=2$ hash functions, meaning that we will need a total of $k\log_2(m)=36$ bits for indexing. We therefore would like our hash function to emit 36 bits, resulting in our SHA3 family bit length selection of SHAKE128(36).

\begin{table}[b]
\centering
\caption{Top 5 $(m,n,k,z)$ with lowest $z$\\
for $f\leq2^{-14}$ and $n\geq1000$}
\label{tbl:optimals}
\begin{tabular}{@{}llllc@{}}
\toprule
z    & m   & n    & k & \multicolumn{1}{l}{bytes per item} \\ \midrule
1977 & $2^{24}$ & 1024 & 1 & 1.931                              \\
2160 & $2^{18}$ & 1028 & 2 & 2.101                              \\
2430 & $2^{15}$ & 1015 & 5 & 2.394                              \\
2943 & $2^{15}$ & 1207 & 6 & 2.438                              \\
3531 & $2^{16}$ & 1516 & 4 & 2.329                              \\ \bottomrule
\end{tabular}
\end{table}

\subsection{Compression Algorithm}

\begin{figure}[t]
    \centering
    \includegraphics[width=\linewidth]{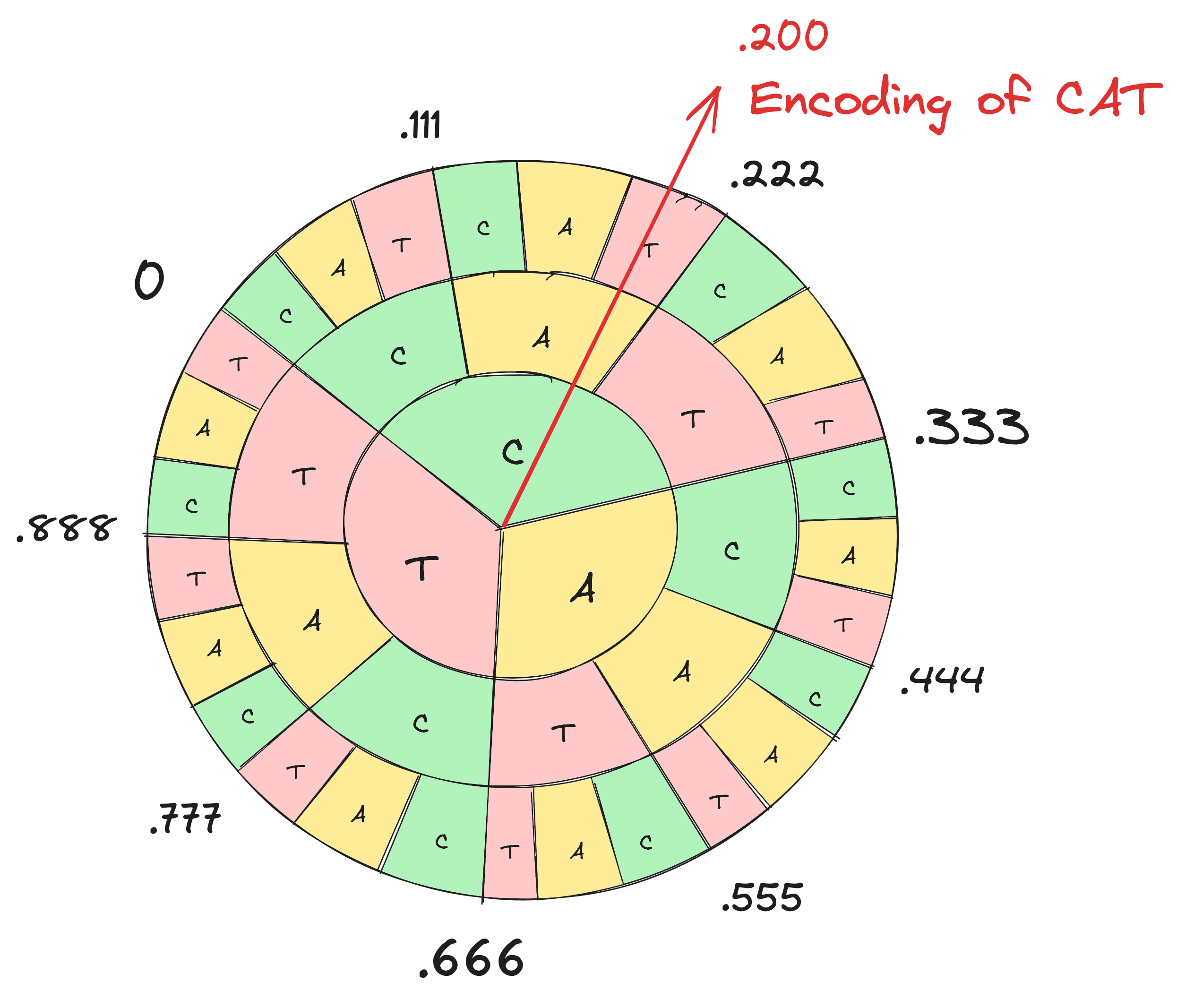}
    \caption{A visualization of compressing the string CAT using arithmetic coding.}
    \label{fig:arithmetic}
\end{figure}

Our Bloom filter parameters were selected to minimize the compressed filter size in addition to the optimizing false positive rates. In \Cref{alg:entropy} we based our optimization on an optimal compressor which could compress to the limit of the entropy function.

In practice, we will have to choose a specific compression algorithm. Following the recommendation of the Compressed Bloom filter proposal~\cite{mitz_compressed_2001}, we select arithmetic coding~\cite{arithmetic_coding_1998}. Arithmetic coding offers near-optimal compression to which the output size approaches the entropy function.

Arithmetic coding works by encoding any bit string as a single arbitrary-precision number in [0,1). The frequency of each symbol in the bit string's alphabet is used to divide the output range into proportional segments. This division is then repeated recursively within each sub-segment. Encoding and decoding are then simply identifying a number that falls within the range of the bit string's location in the nested sub-segment of symbol ranges. We visualize arithmetic encoding in \Cref{fig:arithmetic}.

\subsection{Binary Protocol}

ABOMs are embedded within the binaries emitted by compilers. To accomplish this, we will need to establish a binary protocol for efficiently representing Bloom filters on disk.

To assist with identifying ABOM data, we will begin the binary format with a magic word and a protocol version. We will also need a field to describe the the size of the payload to assist with reading as this length will be variable.

The Bloom filters will be included as a compressed binary payload. $m$ and $k$ will not need to be encoded as they are standardized to $m=2^{18}, k=2$ for all ABOM operations. If multiple Bloom filters are necessary, the bit arrays $A$ will be concatenated in the order of creation prior to compression. The number of Bloom filters $a$ will be packaged in the binary protocol to simplify buffer allocation in deserialization implementations.

Since the Bloom filter bit arrays use a binary alphabet, there are only two symbols for which frequency must be specified to perform optimal arithmetic coding for compression: 0 and 1. However, these frequencies are complementary, so we opt only to include the value of $p(1)$ in the binary protocol. $p(0)$ can be trivially calculated as $1-p(1)$. Since we know that $p(1)$ will be in [0,1], it is inefficient to serialize this value as a floating point. Instead, we choose to serialize $p(1)$ as an unsigned integer of $p(1)$ times the maximum representable numeric value.

Finally, we specify that the entire ABOM binary protocol will be written in little-endian order.

\begin{algorithm}[b]
\caption{ABOM Binary Protocol}
\begin{algorithmic}
\label{alg:binary_protocol}
\STATE \textbf{Header}:
\STATE- Magic Word: `ABOM' \COMMENT{char[4]}
\STATE- Protocol Version: `1' \COMMENT{uint8\_t}
\STATE- Number of Bloom Filters: $a$ \COMMENT{uint16\_t}
\STATE- Arithmetic Model as $p(1) \times (2^{32}-1)$ \COMMENT{uint32\_t}
\STATE- Byte Length of Payload \COMMENT{uint32\_t}
\STATE \textbf{Payload}:
\STATE - Arithmetically-Coded Concatenated Bloom Filters
\end{algorithmic}
\end{algorithm}

\section{Evaluation}

In this section, we will describe our implementation of ABOM, and evaluate its performance in real-world compilations.

\subsection{Implementation}

To analyze whether the ABOMs are feasible in practice, we implemented a set of utilities that create and validate ABOMs.

Specifically, we created three utilities: \texttt{abom}, \texttt{abom-check}, and \texttt{abom-hash}. These utilities were written in Python 3.11 and were designed to work on MacOS 14.0. The source code for the implementation is available on GitHub.\footnote{\href{https://github.com/nickboucher/abom}{github.com/nickboucher/abom}}

\texttt{abom} is invoked via the command line as a wrapper around the \texttt{clang} or \texttt{clang++} LLVM compilers. To add an ABOM to an emitted binary, a user simply prepends the compilation command with the command \texttt{abom}. For example, to compile a program called \textit{foo} with an ABOM, a user may invoke the following command:
$$\texttt{abom clang foo.c -o foo}$$
This will produce a binary that is identical to the binary emitted by the compile command alone but with an added binary section containing the ABOM binary protocol, packaging Bloom filters containing the hashes of all source code files constituting \textit{foo} including recursive upstream dependencies. The utility follows format-specific naming conventions for the added binary sections; for MacOS's MachO binary format, this added segment and section are named \texttt{\_\_ABOM,\_\_abom}. For Linux's ELF and Windows' PE binary formats, this section would be named \texttt{.abom}.

For ABOMs to be exhaustively built, all upstream pre-compiled libraries included in the binary must have also been built with ABOMs. Our implementation of \texttt{abom} allows building if some pre-compiled upstream libraries lack ABOMs, but it will output a warning for each dependency lacking an ABOM.

We further note that a robust ABOM implementation must be able to handle object files, archives, and dynamic libraries in addition to executable binaries. Compilations emitting object files and libraries are treated no different from compilations emitting executable binaries: an \textit{abom} section is packaged within the output binary. Archive files, typically stored as \texttt{.ar} files in build systems, are slightly more complicated; these files are effectively many object files bundled together, and rather than having a single ABOM for the archive each embedded object file will contain its own ABOM. However, as archive files are often frequently re-referenced during large builds, it is reasonable for ABOM tooling to optimize by pre-building an ABOM that is the union of all object file ABOMs embedded in the archive file. Our implementation stores this pre-merged ABOM as a an adjacent file to the archive carrying the same name appended with `.abom'. If such an optimization is used, ABOM tooling must ensure to rebuild this pre-compiled ABOM file each time the archive is modified.

\texttt{abom-check} is also invoked on the command line. This tool is used to check whether a binary contains an ABOM with a specific file as a dependency. \texttt{abom-check} would be used by consumers of software products seeking to check whether software contains a specific known-bad source code file. For example, to check whether a specific source code hash is contained in the binary \textit{foo}, a user may invoke the following command:
$$ \texttt{abom-check foo 7f9c2ba4e0} $$
where \texttt{7f9c2ba4e0} is replaced with the SHAKE128(36) source code hash of interest.

For convenience, we also provide \texttt{abom-hash}, a command line utility to generate SHAKE128(36) hashes of source code files. This tool would be used by software producers or antivirus firms to generate the hash of source code files that are known to be compromised. For example, to generate the hash used by ABOM for source code file \textit{foo.c}, a use may invoke the following command:
$$ \texttt{abom-hash foo.c} $$

\begin{figure}[t]
    \centering
    \includegraphics[width=\linewidth]{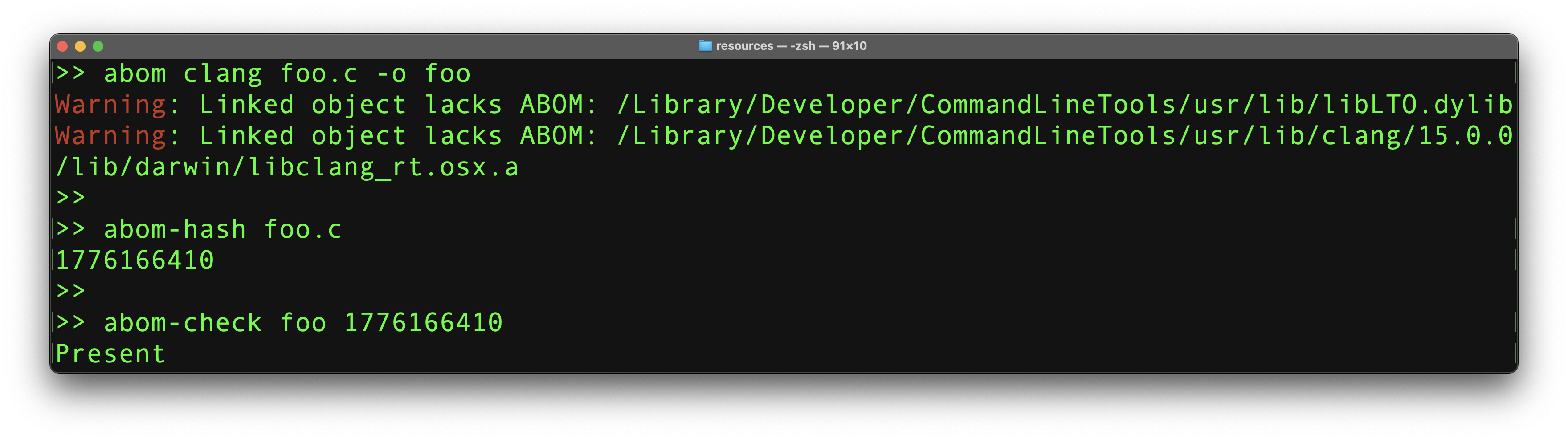}
    \caption{\texttt{abom} and \texttt{abom-check} invoked via the command line on an example program \textit{foo.c}.}
    \label{fig:cli}
\end{figure}

An example invocation is shown in \Cref{fig:cli}. In this example, we first compile the program \textit{foo} with an ABOM packaged in the build. \texttt{abom} outputs two warnings for dynamic libraries lacking upstream ABOMs: this is expected, as the OS-provided libraries aren't built with ABOMs on this system. We next calculate the SHAKE123(36) value of our input file using \texttt{abom-hash}. Finally, we use \texttt{abom-check} to test the output for presence of the previously calculated hash value. The tool correctly outputs that the dependency file is present.

A Python compiler wrapper is not the most efficient way to implement ABOM construction. ABOMs can be built significantly more efficiently within compilers directly; in addition to being implemented in native code, compiler-packaged implementations benefit from not having to reload source code files for hash value calculation. The goal of our ABOM implementation is not performance, but as a proof-of-concept to evaluate ABOM against alternatives, then to bootstrap adoption, and finally as a portable reference implementation for compiler maintainers.

\subsection{Building OpenSSL}

We evaluated the performance of our ABOM implementation against building OpenSSL~\cite{openssl}. We selected OpenSSL because it is a common piece of native C software used broadly across the ecosystem. It is also a reasonably large project with 3,133 C, C++, and ASM source code files in the repository at the time of our experiments. All builds took place on a 2018 MacBook Pro running MacOS 13.

The primary binary emitted from the build process, \texttt{openssl}, contained an ABOM of 992 bytes in size. The binary without the ABOM was 885,764 bytes, so the addition of the ABOM caused a binary size increase of +0.11\%.

The build time with ABOM compilation enabled was 1869 seconds of wall-clock time. Without ABOM compilation, the build took 635 seconds of wall clock time, meaning that the addition of the ABOM added +194\% compilation time. This is a real overhead, but our reference implementation was not designed for performance. Because it not built into the compiler directly, all file reads must be done an extra time by the ABOM tooling. Furthermore, ABOMs must be written to temporary files before injection into compiled binaries. These I/O-heavy tasks would vanish with direct compiler integration, as no additional I/O would be necessary.

\subsection{Building cURL}

We continued evaluation of our ABOM implementation by building cURL, a common tool for network data transfer~\cite{curl}. cURL is written in C and widely used across the ecosystem. It also happens to have a dependency on OpenSSL that enables us to test builds incorporating upstream ABOMs.

\texttt{curl}, the primary binary emitted from the build process, contained an ABOM 3,912 bytes in size. The binary without the ABOM was 191,424 bytes, meaning that the addition of the ABOM caused a size increase of +2.04\%. The build time with ABOM compilation was 186 seconds of wall-clock time, while the build without an ABOM was 65 seconds. This means that the addition of the ABOM added +186\% compilation time using the non-optimized reference implementation.

\subsection{Building GNU Core Utilities}

Finally, for robustness, we extended our evaluation to include a larger collection of programs. We chose to build the GNU Core Utilities, a collection of 107 command line programs commonly associated with *nix operating systems~\cite{coreutils}. Also known as coreutils, this collection contains well known tools such as \texttt{ls}, \texttt{cp}, \texttt{cat}, and \texttt{echo}. It also happens to depend on OpenSSL, for which we again leverage our build of OpenSSL containing an ABOM.

The mean ABOM size for each executable binary emitted from building coreutils was 693.4 bytes with a standard deviation of $\sigma = 47.5$ bytes. The mean binary size without the ABOM was 138069.8 bytes ($\sigma = 41927.7$) with an average ABOM-induced binary size increase of 0.53\% ($\sigma = 0.1\%$). Building all coreutils with ABOM compilation took 600 seconds of wall-clock time, while building coreutils without ABOMs took 137 seconds. This means that the addition of ABOMs added +337\% compilation time using the non-optimized reference implementation.
\section{Discussion}

In this section, we will discuss a collection of broader considerations about ABOM and its adoption.

\subsection{Threat Model}

Much like pre-compiled binaries themselves, ABOMs rely on trusting the compiling entity. Just as pre-compiled binaries could contain adversarial logic divergent from the claims of the publisher, a malicious software publisher could choose to omit or add erroneous information to ABOMs. Like other Bills of Materials, our protocol relies on the trustworthiness of each software publisher. A lack of trust in software publishers must be mitigated by building code from source.

It is also possible for a software distributor to maliciously modify an ABOM after it has been built. An adversary may seek to do this so that their version of the software is less likely to be flagged as vulnerable to future attacks. But standard integrity solutions mitigate this threat: if each binary is cryptographically signed by its producer, that signature will prevent adversaries from tampering with its embedded ABOM. In practice, a significant portion of software is already signed. This means that any addition of ABOMs benefits from these integrity mechanism that are already deployed.

The threat model for ABOMs is one in which the goal is to provide an automatic, robust, transitive record of software dependencies for vulnerability detection in a context where correctness depends on either publisher trust plus signatures, or building software from source.

\subsection{Defining Bill of Materials}

Software Bills of Material are quickly becoming widespread, due at least in part to a new US-government requirement for their suppliers~\cite{biden_eo14028_2021}. Standards are beginning to emerge which attempt to define and implement them.

In this paper, we present something that diverges from most SBOM proposals. Whereas SBOMs tend to include significant quantities of human-comprehensible information about dependencies, ABOMs do not. ABOMs take a minimalist approach to dependency enumeration: they enable querying for the presence of a dependency, and nothing else. ABOMs do this using very little disk space, and do not require any human input to assemble. In contrast, SBOMs tend to be larger, require human input, and are not limited to queries. We enumerate these key differences in \Cref{tbl:sbom}.

\begin{table}[t]
\caption{Comparison of key features differences between ABOMs and traditional SBOMs.}
\label{tbl:sbom}
\resizebox{\linewidth}{!}{
\begin{tabular}{@{}ll@{}}
\toprule
\textbf{ABOM}                             & \textbf{SBOM}                            \\ \midrule
Minimal disk space               & Larger disk space               \\
No human inputs                  & Requires human input/validation \\
Supports dependency queries only & Provides dependency lists       \\
Machine readable                 & Human readable                  \\
Packaged within binaries         & Published alongside binaries    \\ \bottomrule
\end{tabular}
}
\end{table}

Is an ABOM an SBOM? We make no claims whether ABOMs meet any specific SBOM regulatory requirements; that will depend on specific implementation details, such as whether the code contains the metadata required by the NTIA standard. However, we encourage the reader to critically consider the purpose of SBOMs. If SBOMs exist primarily to mitigate the risk of software supply chain attacks, we suggest that ABOMs perform this task with very much less overhead. Some ABOM implementations will surely also contain the extra metadata required for US government compliance; allowing also those ABOMs that do not, within a standard for the ABOM component alone, will enable much more rapid adoption. It should be a universal part of the software engineer's toolkit rather than something added at such expense that it is undertaken only for government work. Projects from the Orange Book to BGPSec have taught that minimising the compliance burden will maximise adoption~\cite{SEv3}.

\subsection{Standards Adoption}

In order to gain widespread adoption, it will be necessary for a standard to emerge that unambiguously describes formats, protocols, and algorithms related to ABOM. We hope that this paper will be the basis for such a standard.

\subsection{Compiler Implementations}

As previously described, ABOM generation is significantly more efficient if implemented within a compiler. Doing so minimizes the I/O required for enumerating and hashing dependencies, and also ensures that binaries need not be modified after they are initially created. There are many compilers in use across the ecosystem, and widespread adoption of ABOM will require a sufficient number of such compilers to add this functionality.

We believe that the most effective way to build robust dependency enumeration across the ecosystem is for mainstream compilers to enable ABOM generation by default. ABOM generation requires no human input; much like other default-enabled compiler security features such as stack canaries, ABOMs have the potential to be something from which all software producers and consumers benefit without their direct knowledge.

\subsection{Towards an AIBOM?}

As machine-learning methods and architectures improve, many software products and services are incorporating some form of machine learning in addition to traditional programming. While ABOMs will work equally well for the training and inference implementations as it will for traditional software, they may also be of use in recording training data. If training data inputs were hashed and inserted into Bloom filters at training time, the resulting ABOMs may provide a space-efficient way to determine whether a specific data point was used to train a model without needing to retain the entire training set. We leave this as an open line of future research.
\section{Related Work}

In this section, we will discuss prior work related to ABOMs and software supply chain attack mitigation.

\subsection{Binary-Embedded Metadata}

As described in \Cref{sbom}, SBOMs are the traditional solution for representing software dependencies. However, SBOMs suffer from being large in size, requiring human input, and necessitating the distribution of auxiliary files with software.

The earliest proposal to embed dependency information within binaries that we could find was a rejected 2021 Fedora Linux proposal~\cite{fedora_f35} to embed package names within ELF object files~\cite{lwn_elf}. This proposal, although not adopted, aimed to assist with debugging by placing package name information in a location that would be visible in core dumps. While this proposal was not related to supply chain attacks, it introduces the idea of embedding package information within compiled binaries.

\subsection{OmniBOR}

Later work suggesting that binary-embedded metadata could be used to used to perform run-time vulnerability detection resulted in a project known as OmniBOR~\cite{omnibor_whitepaper}. OmniBOR uses the hashes created by git, known as gitoids, to construct a Merkle tree of the source code dependency graph. The root hash of this tree is then embedded into compiled binaries.

OmniBOR provided a step forward in proposing that vulnerabilities could be identified via binary-embedded information, but has the major shortcoming that OmniBOR's embedded dependency information is not self-contained. By including only the root of the Merkle tree, it is not possible to generically identify whether a certain dependency is contained within the tree without also providing a reconstruction of the tree out of band. While it would be possible to extend this proposal to include entire dependency trees, the disk space requirements and dependency query times would quickly rise with the number of dependencies.

ABOM offers an alternative proposal that is self-contained, highly space efficient, and offers near-constant time dependency queries in practice.
\section{Conclusion}

We propose the Automatic Bill of Materials, or ABOM, a novel technique to embed dependency information in compiled binaries. This information enables the detection of supply chain attacks by offering the ability to directly query binaries for the presence of specific source code files. ABOMs represent dependencies as hashes stored in compressed Bloom filters. This protocol allows ultra space efficient dependency storage by allowing for a low probability of false positives. ABOMs are fully automated and can be enabled in compilers without any developer intervention, in the same way as memory-safety mitigations such as stack canaries. ABOMs provide a zero-touch, backwards-compatible, language-agnostic tool that can quickly identify software supply chain attacks and in turn bolster the security of the entire software ecosystem.

\bibliographystyle{IEEEtran}
{\small
\bibliography{sources}}

\end{document}